# Isothermal titration calorimetry as a powerful tool to quantify and better understand agglomeration mechanisms during interaction processes between TiO$_2$ nanoparticles and humic acids

Frédéric Loosli [a], Letícia Vitorazi [b], Jean-François Berret [b] and Serge Stoll [a,*]

**Affiliations:**

[a] *Group of Environmental Physical Chemistry, University of Geneva, F.-A. Forel Institute Section des Sciences de la Terre et de l'Environnement, 10 route de Suisse, 1290 Versoix, Switzerland*
[b] *Laboratoire Matière et Systèmes Complexes, UMR 7057 Université Paris-Diderot/CNRS, Bâtiment Condorcet, 10 rue Alice Domon et Léonie Duquet, F-75205 Paris cedex 13, France*

**Abstract**
The association processes between engineered TiO$_2$ nanoparticles and Suwannee River humic acids are investigated by isothermal titration calorimetry and by measuring the exchanged heat during binding process allowing the determination of thermodynamic (change of enthalpy, Gibbs free energy and entropy) and reaction (binding affinity constant, reaction stoichiometry) parameters. Our results indicate that strong TiO$_2$-Suwannee River humic acid interactions are entropically and enthalpically favorable with exothermic binding reactions and that the mixing order is also an important parameter. High humic acid concentrations induce homoagglomeration ("self" assembly) and are shown to favor an enthalpically driven association process. Light scattering techniques are also considered to investigate the influence of TiO$_2$ surface charge modifications and agglomeration mechanisms. Patch and bridging mechanisms are found to result into the formation of large agglomerates once charge inversion of TiO$_2$-humic acid complexes is achieved. Moreover van der Waals interactions are also found to play a significant role during interaction processes due to the amphiphilic character of humic acids.

[*]**Corresponding Author and Address:**
Serge Stoll [*]: Group of Environmental Physical Chemistry, University of Geneva, F.-A. Forel Institute, Section des Sciences de la Terre et de l'Environnement, 10 route de Suisse, 1290 Versoix, Switzerland
Phone: + 41 22 379 0333; Fax: + 41 22 379 0302 ; email: serge.stoll@unige.ch

## 1. Introduction

A better understanding on the fate and behavior of engineered nanoparticles (ENPs) in the presence of aquagenic compounds is of great importance for the risk assessment associated to ENPs entering environmental aquatic systems [1-5]. Indeed nanomaterials are produced in large and growing amount [6] due to their very unique electronic and surface chemistry properties and huge specific surface area [7]. ENPs are then expected to enter aquatic natural systems through surface runoff, accidental discharge but also due to the lack of efficiently in removing them in wastewater treatment plants [8-10]. Once in aquatic systems ENP stability is strongly influenced by water physicochemical properties, i. e. pH, ionic strength [11-13], ENP intrinsic properties, i.e. size, shape, surface charge and chemistry [14-15], and presence of aquagenic compounds such as natural colloids and living microorganisms [16-17]. Complex formation between ENPs and natural compounds strongly modifies the ENP stability, fate, transport, bioavailability and effect towards living organisms [18-19].

An important class of organic colloids is represented by natural organic matter (NOM). The larger fraction of NOM is composed by humic substances, with up to 30-50% of surface water organic matter [20]. Humic substances are derived from plant and animal residues through humification processes [21]. Interaction processes between ENPs and the soluble fraction of humic substances (Fulvic and humic acids) have been investigated in many studies [22-24]. Their presence was shown to deeply modify the ENP surface charge and resulting stability through electrostatic interactions and steric effects once adsorbed on ENPs [25-27]. Presence of NOM was not only found to promote ENP agglomeration or stabilization, but also to induce the partial fragmentation of already formed ENP agglomerates [28-30]. Agglomeration versus fragmentation was found dependent on the NOM nature, ENP surface properties and concentration ratio between them.

Most of these studies investigated the ENP stability, for different experimental conditions, by determining the ENP surface charge modification and the resulting state of agglomeration (size and fractal dimension) to understand the influence of pH, NOM properties, electrolyte concentration and valency. In the present study we focus on a different but important complementary aspect related to the quantification of the energies associated to the interaction processes and agglomeration mechanisms. Isothermal titration calorimetry (ITC) is used here to quantify the complexation between $TiO_2$ ENPs and humic acids. ITC is an instrumental technique which permits to determine, in a single experiment, all interaction thermodynamic parameters ($\Delta H$, $\Delta S$ and $\Delta G$) and to provide information on the reaction stoichiometry and binding affinity. ITC has been applied to study binding reactions for the self-assembly of supramolecular polymers and protein-substrate interactions [31-34]. In this study we are using a novel approach, via ITC measurements, to get a quantitative insight of the interaction energies between $TiO_2$ ENPs and humic acids. $TiO_2$ ENPs are one of the most produced nanomaterials [6, 35-36] as being used in many industrial domains such a in the food, cosmetic and painting industries [37-39]. $TiO_2$ ENPs are likely to be already present in the natural aquatic systems in the ng $L^{-1}$ to µg $L^{-1}$ range [9, 40]. The influence of NOM and water properties on their stability thoroughly investigated [41-42].

In the present study all thermodynamic reaction parameters associated to the complexation processes are considered together including ENP surface charge modification and size evolution to propose a detailed quantification of the energy involved during interaction processes. Comparison is also made with dynamic light scattering and electrophoretic mobility measurements to a better description agglomeration mechanisms (patch and bridging) between $TiO_2$ ENPs and humic acids.

## 2. Materials and methods

### 2.1. Materials

A 5 g $L^{-1}$ $TiO_2$ dispersion was prepared by dilution of a 15% wt $TiO_2$ suspension, after homogeinization, obtained from Nanostructured & Amorphous Material Inc (Houston, TX,

USA) with Milli Q water (Millipore, Zoug, ZG, Switzerland, with R >18 MΩ.cm, T.O.C. <2 ppb). NaOH and HCl (1 M, Titrisol®, Merck, Zoug, ZG, Switzerland) were used after dilution to adjust the ENP dispersions at pH 3.8 and 10.4. The $TiO_2$ dispersion was then dialyzed against water (pH 3.8 and 10.4) with a 12-14 kDa cutoff dialysis membranes (Spectrum Laboratories, Inc., Rancho Dominguez, CA, USA). The dialyze solvant was first used to prepare humic acid (Suwannee River humic acids (SRHA), Standard II, International Humic Substances Society, Denver, CO, USA) solutions with concentrations equal to 1.25 mM in term of charge concentration (600 mg $L^{-1}$ for pH 3.8 and 201 mg $L^{-1}$ for pH 10.4) which were stirred overnight. Dialysis was realized to minimize the change of enthalpy due to the dilution processes during titration. The rather low molecular weight of SRHA [43] does not allow their dialyses. The water from dialysis was also used to dilute the $TiO_2$ and SRHA suspensions to experimental concentrations (from 0.1 to 3.5 g $L^{-1}$ for $TiO_2$ and from 0.0375 to 0.75 mM for SRHA).

## 2.2. Isothermal titration calorimetry measurement

A VP-ITC calorimeter (MicroCal Inc., Northampton, MA, USA) with a sample cell volume equal to 1.4643 mL was used to determine the heat exchange between $TiO_2$ ENPs and SRHA. After a preliminary injection of 2 μL of the first compound (ligand (L)), 28 successive injections of 10 μL into the sample cell, containing the other compound of interest (Macromolecule (M)), were realized with injection duration equal to 20 s and a delay between two successive additions of 300 s. The stirring speed was set to 307 rpm and the working temperature to 298.15 K for all experiments. The plot of the heat of exchange ($dQ/dn_L$) as a function of the molar charge ratio (Z = [L]/[M], where [L] and [M] are the ligand, and macromolecule molar charge concentration, respectively) is fitted with a Multiple Non-Interacting Sites (MNIS) model (Eq. 1) where the sites do not exhibit cooperative binding behavior [44]. In order to determine the fitting parameters, the following equation is used.

$$\frac{dQ}{dn_L}(Z) = \frac{1}{2} \Delta H_b \left[ 1 + \left(1 - \frac{[L]}{n[M]} - \frac{1}{n K_b [M]}\right) \left(\left(1 + \frac{[L]}{n[M]} + \frac{1}{n K_b [M]}\right)^2 - \frac{4[L]}{n[M]}\right)^{-\frac{1}{2}} \right] \quad (1)$$

In this equation, the fitting parameters are $\Delta H_b$ [kJ $mol^{-1}$], $K_b$ [$M^{-1}$] and n, and they represent the binding enthalpy (energy involved during association processes), affinity binding constant (affinity between the two compounds) and reaction stoichiometry, respectively. These parameters are adjusted to fit the experimental curves with the mathematical MNIS model given by Eq.1. Then, the Gibbs free energy, $\Delta G$ [kJ $mol^{-1}$], and the change of entropy, $\Delta S$ [kJ $K^{-1}$ $mol^{-1}$], are calculated from the fitting parameters with $\Delta G = -RT \ln K_b$ and $\Delta S = (\Delta H - \Delta G)/T$. The ligand and macromolecule charge numbers for the ith-injection are equal to:

$$L_i = L_{i-1} + V_i \cdot [L]_{m,} \cdot N_A \quad and \quad M = [M]_m \cdot V_{cell} \cdot N_A \quad (2)$$

$$with \ [TiO_2]_m = [TiO_2]_M \cdot S_A \cdot \sigma_{TiO_2} \cdot 10^{18} \cdot \frac{1}{N_A} \quad (3)$$

$$and \ [SRHA]_m = [SRHA]_M \cdot Q_{tot} \cdot 10^{-3} \cdot \frac{C\%}{100} \quad (4)$$

$$where \ Q_{tot} = \left(\frac{Q_1}{1+\left(K_1[H^+]\right)^{\frac{1}{n_1}}}\right) + \left(\frac{Q_2}{1+\left(K_2[H^+]\right)^{\frac{1}{n_2}}}\right) \quad (5)$$

In Eq. (2), $V_i$ and $V_{cell}$ represent the volume of ligand injected and the cell volume respectively. [L or M]$_m$ corresponds to the mol of charge of ligand or macromolecule per unit of volume and $N_A$ to the Avogadro constant. The concentration in term of mol of charge per unit of volume for TiO$_2$ and SRHA are expressed in Eq. (3) and (4) respectively. In Eq. (3), [TiO$_2$]$_M$ represents the TiO$_2$ mass concentration, $S_A$ the ENP specific surface area [m$^2$ g$^{-1}$] and $\sigma_{TiO_2}$ the TiO$_2$ hydroxyl sites density [sites nm$^{-2}$]. In Eq. (4), [SRHA]$_M$ represents the SRHA mass concentration, $Q_{tot}$ the SRHA overall charge density [meq g C$^{-1}$] and C% the SRHA percent of carbon, which is equal to 52.63% [45]. For the modified Henderson-Hasselbalch equation (Eq. 5), the values of maximum charge densities $Q_1$ and $Q_2$ of SRHA carboxylic and phenolic binding sites, the two dissociation constants $K_1$ and $K_2$ and empirical parameters $n_1$ and $n_2$ were taken from Ritchie et al. study dealing with the proton binding of standard SRHA [46].

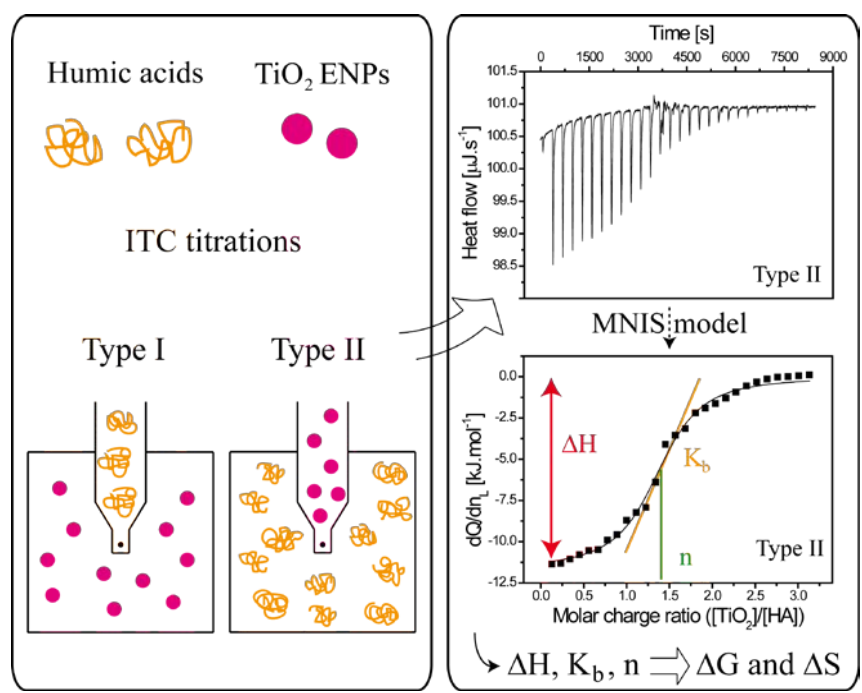

***Fig. 1*** *- ITC type I (SRHA in TiO$_2$ dispersion) and type II (TiO$_2$ in SRHA) titrations. ITC measurements give the heat flow for each of the 28 injections and after fitting the integrated data with the multiple non-interacting sites (MNIS) model the enthalpy of exchange ΔH, the binding constant $K_b$ and the reaction stoichiometry n are determined. It allows the calculation of the entropy ΔS and total free energy ΔG for TiO$_2$-SRHA interaction.*

TiO$_2$ charge concentrations were estimated based on manufacturer data (primary TiO$_2$ diameter equal to 15 nm and in good agreement with the mode value of TiO$_2$ number size distribution (ESI† Fig. S1)) to calculate the $S_A$ and on Kominami et al. study to estimate $\sigma_{TiO_2}$ [47]. A $S_A$ equal to 100 m$^2$ g$^{-1}$ and 5 sites nm$^{-2}$ $\sigma_{TiO_2}$ were used to determine the TiO$_2$ charge concentration and the factor of conversion between the mass and charge concentration was set so as a 1 g L$^{-1}$ TiO$_2$ dispersion corresponds to a 0.83 mM charge concentration.

The mixing order of the two compounds, which is an important issue to consider, was investigated to better understand the TiO$_2$-SRHA and SRHA-TiO$_2$ interactions and agglomeration process as illustrated in Fig. 1. In a first set of experiments (type I) SRHA was playing the role of ligand and was added to the $TiO_2$ dispersions. In the second set of experiments (type II) $TiO_2$ ENPs (L) were added to SRHA (M).

Experiments were made at pH 3.8 (and at pH 10.4) without addition of electrolyte. Such a pH value was used to address the interaction between isolated, dispersed ENPs and SRHA. It is important to note that results remains valid as long as pH < $pH_{PCN,TiO_2}$. The domain of concentration investigated was from 0.25 mM SRHA in 0.1 g $L^{-1}$ $TiO_2$ to 1.25 mM SRHA in 0.5 g $L^{-1}$ $TiO_2$ for Type I experiments and from 0.7 g $L^{-1}$ $TiO_2$ in 0.0375 mM SRHA to 3.5 g $L^{-1}$ $TiO_2$ in 0.1875 mM alginate for Type II experiments. Such $TiO_2$ concentration are higher than the expected environmental concentration, which are in the ng to μg $L^{-1}$ range [8-10], but necessary to obtain an optimum signal with the calorimeter.

## 2.3. Zeta potential and size distribution measurements

Zeta (ζ) potential values and z-average hydrodynamic diameters of $TiO_2$ and SRHA suspensions as a function of pH as well as $TiO_2$ in presence of SRHA as a function of charge ratio were determined by laser Doppler velocimetry and dynamic light scattering (Zetasizer Nano ZS instrument, Malvern Instruments, Worcestershire, UK). The instrument was operating at 298.15 K. For ζ potential values determination the Smoluchowski approximation model was applied according to the formation and presence of large agglomerates [30]. All polydispersity indexes were found below 0.6.

# 3. Results and discussion

ITC experiments were realized to determine the binding properties mainly at pH < $pH_{PCN,TiO_2}$. Such a pH domain favors electrostatic interactions. $TiO_2$ ENPs are positively charged (ζ potential = +40.9 ± 1.4 mV (mean ± standard deviation on mean of triplicates) whereas SRHA are negatively charged (ζ potential = -37.2 ± 1.9 mV) as shown in Fig. 2 where the ζ potential is represented as a function of pH for both compounds. Some experiments were also made at pH > $pH_{PCN,TiO_2}$, where both compounds exhibit the same charge (negatively charged $TiO_2$ and SRHA) to check if only steric and electrostatic repulsions are involved or if other forces such as van der Waals interactions are also expected to play a significant role during the interaction processes. At pH < $pH_{PCN,TiO_2}$ and pH > $pH_{PCN,TiO_2}$ $TiO_2$ ENPs are dispersed as shown in Fig. S1† with z-average diameter equal to 50 nm whereas the SRHA z-average diameter is found constant with pH changes and equal to 379 ± 19 nm (Fig. S2).†

## 3.1. $TiO_2$-SRHA thermodynamic and reaction binding parameters determined by ITC

### 3.1.1. Type I titration - Titrations of $TiO_2$ ENP dispersions by SRHA

At pH < $pH_{PCN,TiO_2}$, when SRHA are added to $TiO_2$, the interactions are found important as shown in the real time thermogram in Fig. 3a where the titration process for a 0.1 g $L^{-1}$ $TiO_2$ dispersion with a 0.25 mM SRHA is addressed. Negative peaks in the thermogram indicates that the interaction between $TiO_2$ and SRHA is an exothermic process since the thermogram represents the power generated by the calorimeter along the titration to maintain a small and constant difference of temperature between a reference cell (fill with the reaction solvent: water) and the reaction cell, both being located in an adiabatic jacket. After the first injection, corresponding to the peak of smaller intensity (as the injected volume is here equal to 2 μL of SRHA instead of the 10 μL "conventional" injection volumes), the next fifteen peak intensities (each peak is referring to a single injection) are of similar value. This denotes that $TiO_2$ in such condition free surface sites are available for SRHA adsorption. Then, as the titration is progressing, the peak intensities are decreasing due to the restriction of $TiO_2$ binding site available for further SRHA adsorption until site saturation is reached.

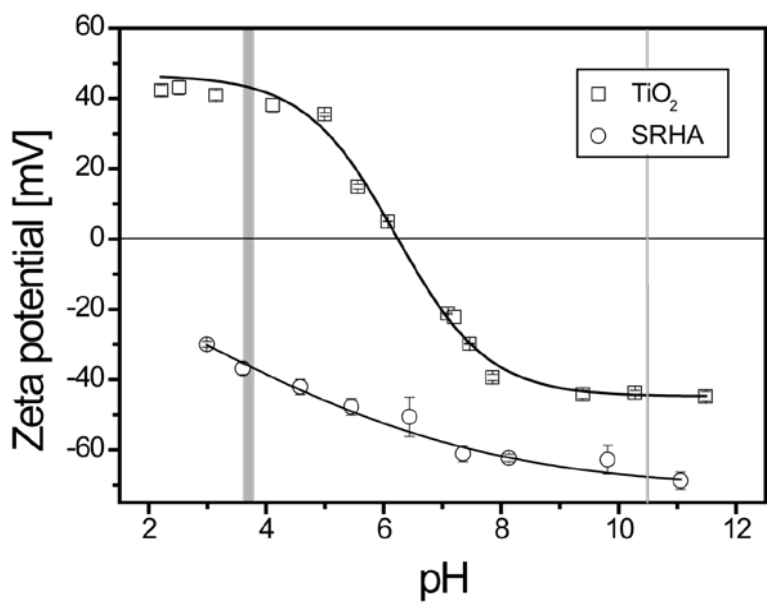

*Fig. 2 -* Zeta potential values of $TiO_2$ (open squares) and SRHA (open circles) as a function of pH. $TiO_2$ isoelectric point is found here equal to 6.2 ± 0.1 whereas SRHA exhibit a negative structural charge in the full pH range. At pH 3.1 $TiO_2$ and alginate have ζ potential values equal to +40.9 ± 1.4 mV and -37.2 ± 1.9 mV respectively (large gray vertical line). At pH 10.4 both compounds are negatively charged (narrow gray vertical line). $[TiO_2]$ = 50 mg $L^{-1}$, $[SRHA]$ = 100 mg $L^{-1}$ and $[NaCl]$ = 0.001 M.

Then mainly dilution effect is observed as shown by the low heat flow recorded by the calorimeter. Fig. 3b represents the energy of exchange per mol of injectant ($dQ/dn_L$) as a function of SRHA over $TiO_2$ charge ratio (Z = [L]/[M]) and is obtained by integration of the previous thermogram. The plot is then fitted with the MNIS model to determine the thermodynamic and reaction parameters associated to the interaction process [44]. The interaction process, for the titration of a 0.1 g $L^{-1}$ $TiO_2$ dispersion by a SRHA 0.25 mM solution, is exothermic as $\Delta H_b$ is equal to -18.3 kJ $mol^{-1}$. The binding affinity between the ENP and SRHA, which is expressed by $K_b$, is equal to 4.3 × $10^6$ $M^{-1}$. The reaction stoichiometry is found equal to 0.41. The fitting parameters then permitted the calculation of $\Delta G$ and $\Delta S$ which are equal to -37.9 kJ $mol^{-1}$ and 65.8 J $K^{-1}$ $mol^{-1}$, respectively. Three other type I experiments were done at different concentrations, but by keeping the ratio between $TiO_2$ and SRHA concentrations constant in order to evaluate the influence of relative concentration on the interaction processes. The real time thermograms and the plots of the heat exchange as a function of SRHA over $TiO_2$ charge ratio for these experiments are presented in Figs. S3 to S5.† All determined and calculated thermodynamic and reaction parameters are represented in Table 1. To clearly see the main driving force (enthalpy versus entropy) the values of T $\Delta S$ are given in this table.

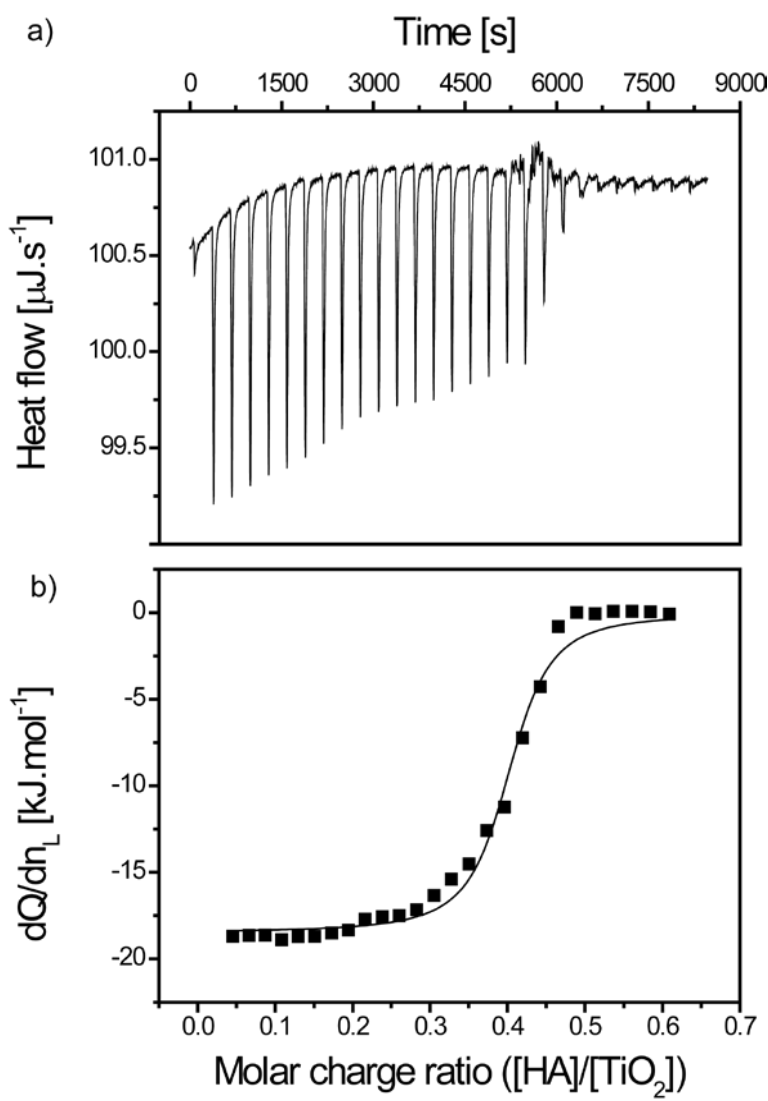

***Fig. 3*** - *a) Real-time thermogram for $TiO_2$ 0.1 g $L^{-1}$ titration with SRHA 0.25 mM at pH < $pH_{PCN,TiO_2}$ and at 298.15 K. The heat flow refers to the thermal compensation of the calorimeter to keep the sample at a constant temperature. Here negative peaks indicate an exothermic reaction. After about twenty injections sites saturation occurs and only weak binding energy is observed. b) The respective integrated heat data ($dQ/dn_L$) as a function of molar charge ratio ([SRHA]/[$TiO_2$]) is fitted with the multiple non-interacting sites (MNIS) model. The binding enthalpy, the binding constant and the reaction stoichiometry where found here equal to -18.3 kJ $mol^{-1}$, 4.3 × $10^6$ $M^{-1}$ and 0.41, respectively.*

$TiO_2$ titration with SRHA are thus spontaneous reactions with high Gibbs energy value ($\Delta G$ <-30 kJ $mol^{-1}$). The interaction energies are found to lead to the formation of $TiO_2$-SRHA complexes due to favorable enthalpy conditions ($\Delta H_b$ <0) but also to an entropic gain (T $\Delta S$ >0). Relative concentration is found to influence the binding energy as higher concentration involved higher $\Delta H_b$ values. This behavior can be attributed to the importance of SRHA homoagglomeration ("self" assembly) due to agglomerates weakly bounded by hydrophobic interactions and H-bonding [48-49] which is concentration dependent [50]. Indeed larger SRHA agglomerates are expected to promote interaction with an increasing amount of $TiO_2$ ENPs in comparison to smaller SRHA agglomerates. The decrease of the binding affinity, and thus calculated Gibbs free energy, with the increase of the relative experimental concentrations is due to the relation between $K_b$ and c according to $K_b \sim c^{-2}$ [44]. The reaction stoichiometry is found, for type I titrations, to slightly decrease. Overall it denotes that $TiO_2$ ENPs are not fully coated with SRHA which is in agreement with the SRHA structure. Indeed humic acids are often considered as heterogeneous semi-rigid globular macromolecules [51-53]. The entropy gain is lower when relative concentrations are increasing. This is not only due to the decrease of the Gibbs free energy for higher concentration but also to the fact that for larger SRHA homoagglomerates the gain of entropy is smaller owing to the lower conformational entropy gain that occurs during binding process and lower gain of entropy due to water molecules release. The complex formation is therefore mainly driven by the importance of the energy of binding ($\Delta H_b < -T \Delta S$) excepted for the lowest relative concentration investigated in this study for which $\Delta H_b$ and $-T \Delta S$ have similar values.

Table 1: Fitting parameters $\Delta H_b$, $K_b$ and n from ITC analysis of the integrated heats with MNIS model and calculated $\Delta G$ and $T \Delta S$ (from $K_b$ and $\Delta H_b$ values) for type I titration.

| SRHA in $TiO_2$ | $\Delta H_b$ [kJ mol$^{-1}$] | $K_b$ [M$^{-1}$] | n | $\Delta G$ [kJ mol$^{-1}$] | $T \Delta S$ [kJ mol$^{-1}$] |
|---|---|---|---|---|---|
| 0.25 mM in 0.1 g L$^{-1}$ | -18.3 | $4.3 \times 10^6$ | 0.41 | -37.9 | 19.6 |
| 0.50 mM in 0.2 g L$^{-1}$ | -22.9 | $2.5 \times 10^6$ | 0.31 | -36.5 | 13.6 |
| 0.75 mM in 0.3 g L$^{-1}$ | -23.2 | $1.1 \times 10^6$ | 0.34 | -34.6 | 11.4 |
| 1.25 mM in 0.5 g L$^{-1}$ | -24.4 | $8.5 \times 10^5$ | 0.32 | -33.9 | 9.4 |

### 3.1.2. Type II titration - Additions of $TiO_2$ in SRHA solutions

where the real-time thermogram and the respective integrated heat of exchange per mol of $TiO_2$ as a function of molar charge ratio for a 0.7 g L$^{-1}$ $TiO_2$ in 0.0375 mM SRHA are represented. When SRHA is titrated by $TiO_2$ ENPs at pH < $pH_{PCN,TiO_2}$, the interaction process is also entropically and enthalpically favorable. Indeed $\Delta H_b$ and $T \Delta S$ are equal to -11.7 kJ mol$^{-1}$ and 20.9 kJ mol$^{-1}$, respectively. A binding constant of $5.1 \times 10^5$ in the more diluted conditions and a 1.41 reaction stoichiometry suggest an important binding affinity and non fully coated $TiO_2$ ENPs. Experiments at different concentration are also realized and all parameters represented in Table 2. Real time thermograms and respective heat exchange plots are represented in S6 to S8.†

The reaction stoichiometry is not dependent on the relative concentration investigated (n = 1.41 ± 0.15). Moreover the enthalpy of binding is not significantly influenced by the increase of concentration (-12.3 ± 0.8 kJ mol$^{-1}$).

When comparison is made between the two titration procedures an important difference in the value of the binding enthalpy is observed due to the much higher SRHA concentration, and thus larger SRHA homoagglomerate formation and higher enthalpy of interaction for type I. When SRHA are added to solution containing $TiO_2$ the enthalpy is significantly more important because of the possibility to complex more ENPs. SRHA homoagglomeration phenomena is also the reason why type I titration are mainly driven by enthalpy whereas for type II interaction processes are driven by an important gain in entropy as the total free Gibbs free energy is similar for both titration types. The entropic gain is arising from the SRHA and ENP counter-ions and water molecules release during adsorption processes. It should be noted that larger loss of entropy is associated with a larger increase in enthalpy as suggested by the enthalpy-entropy compensation [54-55]. Similar total free energy values were observed during the association process between ZnO ENPs with lysozyme, as well as between proteins and amino acid functionalized gold ENPs [56-58]. However in these studies the interactions ($K_b = 0.9 \times 10^6$ M$^{-1}$) were enthalpically favorable but entropically unfavorable due to conformational restriction of proteins.

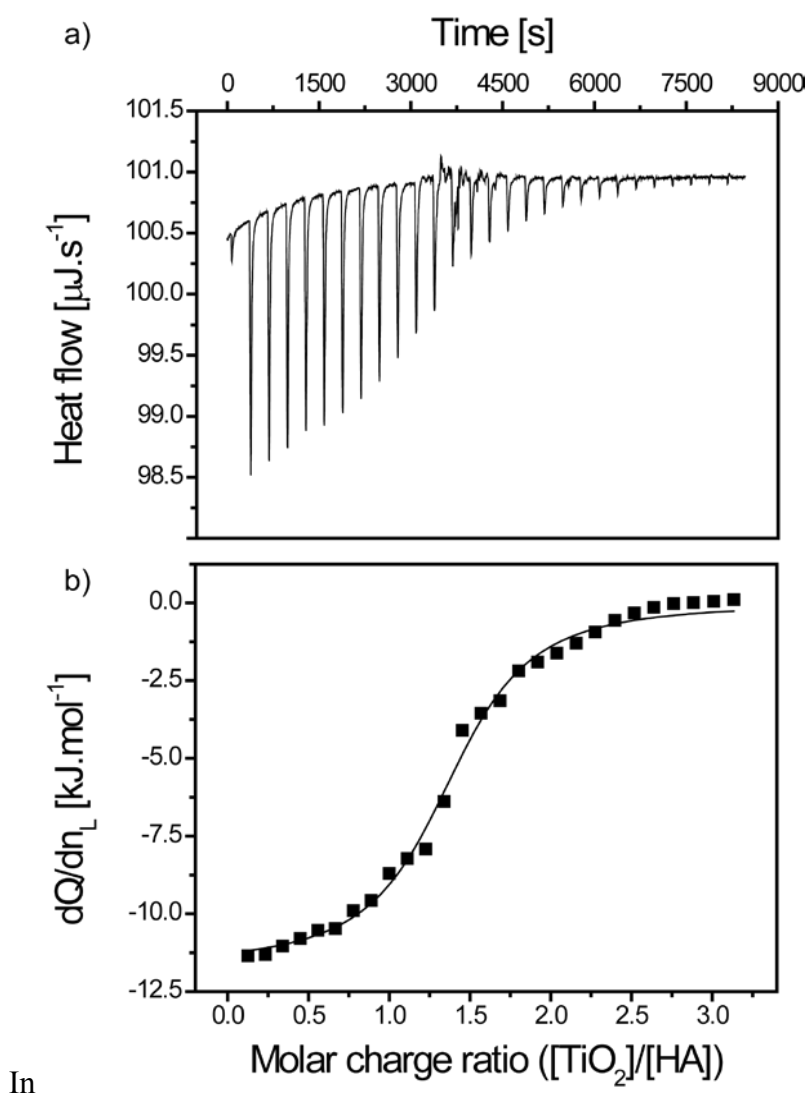

In

*Fig. 4* - a) *Real-time thermogram for SRHA 0.0375 mM titration with TiO$_2$ 0.7 g L$^{-1}$ at pH < pH$_{PCN,TiO_2}$ and at 298.15 K. Negative peaks indicate an exothermic reaction and after about twenty injections sites saturation occurs and only weak binding energy is observed. b) The respective integrated heat data as a function of TiO$_2$ over SRHA molar charge ratio is fitted with the MNIS model. The binding enthalpy, the binding constant and the reaction stoichiometry are found here equal to -11.7 kJ mol$^{-1}$, 5.1 × 10$^5$ M$^{-1}$ and 1.41, respectively.*

Table 2: Fitting parameters $\Delta H_b$, $K_b$ and n from ITC analysis of the integrated heats with MNIS model and calculated $\Delta G$ and $T \Delta S$ (from $K_b$ and $\Delta H_b$ values) for type II titration.

| TiO$_2$ in SRHA | $\Delta H_b$ [kJ mol$^{-1}$] | $K_b$ [M$^{-1}$] | n | $\Delta G$ [kJ mol$^{-1}$] | $T \Delta S$ [kJ mol$^{-1}$] |
|---|---|---|---|---|---|
| 0.7 g L$^{-1}$ in 0.0375 mM | -11.7 | 5.1 × 10$^5$ | 1.41 | -32.6 | 20.9 |
| 1.4 g L$^{-1}$ in 0.0750 mM | -12.0 | 3.3 × 10$^5$ | 1.60 | -31.5 | 19.5 |
| 2.1 g L$^{-1}$ in 0.1125 mM | -12.2 | 2.5 × 10$^5$ | 1.40 | -30.8 | 18.6 |
| 3.5 g L$^{-1}$ in 0.1875 mM | -13.4 | 1.6 × 10$^5$ | 1.24 | -29.6 | 16.2 |

Reaction stoichiometries are not equal to unity even if being in presence of an electrostatic interaction scenario. Indeed as SRHA are relatively heterogeneous in size and highly charged macromolecules in these working conditions, bridging and patch mechanisms are expected to play an important role during the interaction processes [59-61]. Therefore a significant number of ENPs surface sites are hindered due to conformational restrictions and surface charge heterogeneity, respectively, which favors a reaction stoichiometry lower than unity for type I titration, and higher than unity for type II.

Another interesting behavior when investigating the association process between TiO$_2$ and SRHA is the weak exchange energy which is still observed for high [L] over [M] charge ratio. It means that non electrostatic interactions are involved even if EPN surface sites being no longer available. Such interaction energies, which are more important for high SRHA concentrations, can be linked to the amphiphilic character of SRHA that is known to exhibit significant van der Waals interactions [62].

To verify the presence of such van der Waals interactions, the complexation process between TiO$_2$ and SRHA is also investigated at pH 10.4 for a titration of a 5 g L$^{-1}$ TiO$_2$ dispersion by SRHA 1.25 mM. At pH > pH$_{PCN,TiO_2}$ both compounds are negatively charged (Fig. 2). If only electrostatic interactions are involved during the association processes the real time thermograms and the respective heat exchange plots for both the titration (SRHA 1.25 mM in 5 g L$^{-1}$) and for the dilution (SRHA 1.25 mM in water) should be identical (or at least very similar). This is not the case and an association process is shown to happen between TiO$_2$ and SRHA in this unfavorable electrostatic scenario (Figs. S9 and S10)†. Van der Waals interactions are shown to be significant with energy exchange in the early titration stage three times greater than the dilution effect and of the order of few kJ mol$^{-1}$. The presence of van der Waals interactions at pH > pH$_{PCN,TiO_2}$ is also in good agreement with previous study where SRHA were shown to adsorbed onto TiO$_2$ ENPs with decrease of the electrophoretic mobility and increase of ENP size when negatively charge SRHA were added to negatively charged TiO$_2$ [63].

### 3.2. Effect of TiO$_2$-SRHA agglomeration and surface charge on binding heat of exchange

In order to understand the influence of agglomerate formation and surface charge on the heat of exchange between TiO$_2$ and SRHA, electrophoretic mobility measurements and size determination were realized and then comparison was made with the previous heat exchange data plots obtained by ITC. Experiments were realized at pH < pH$_{PCN,TiO_2}$, for 0.25 mM

SRHA in 0.1 g L$^{-1}$ TiO$_2$ (Type I) and 0.7 g L$^{-1}$ in 0.0375 mM SRHA (Type II) with a 300 s delay between each successive titrant addition (identical than for ITC experiments). As this delay time was not long enough to undergo both ζ potential and z-average diameter determination, triplicates for each of these parameters were measured separately.

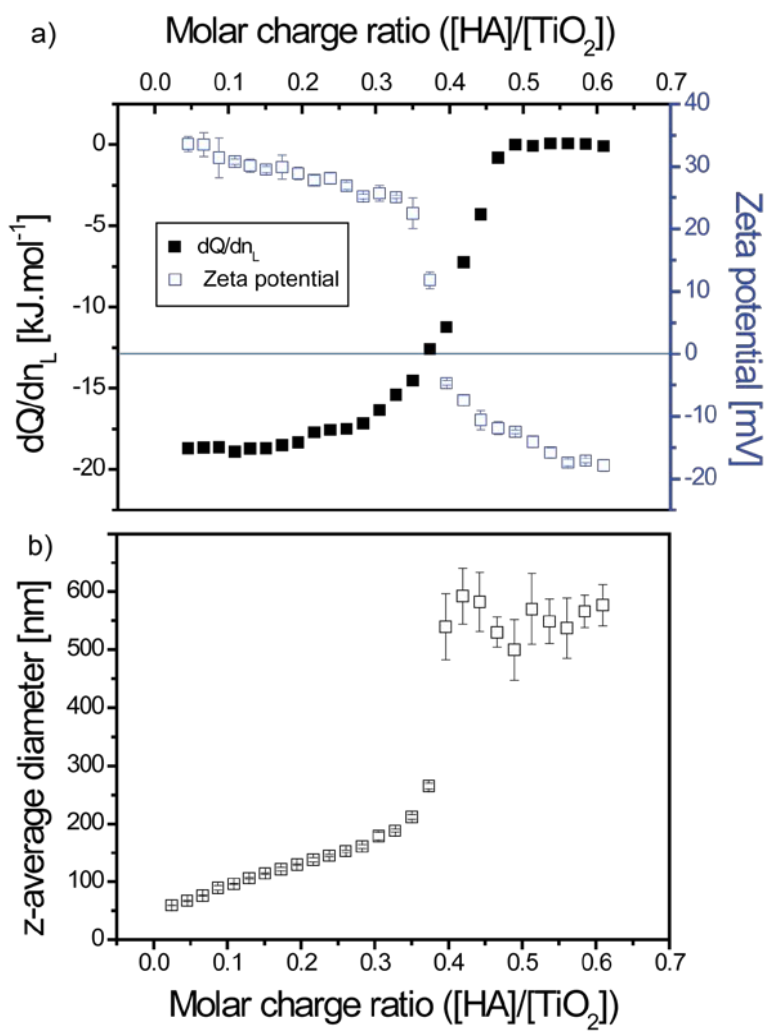

***Fig. 5*** - *a) Integrated heat data and ζ potential values as a function of SRHA over TiO$_2$ charge ratio for TiO$_2$ 0.1 g L$^{-1}$ titration with SRHA 0.25 mM at pH < pH$_{PCN,TiO_2}$. For a ratio up to 0.35, the binding enthalpy and the ζ potential values are slightly decreasing. Then charge inversion (ζ potential = -4.7 ± 0.4 mV) is observed for Z = 0.40 and for Z >0.48 site saturation occurs and weak and constant interaction are observed. b) z-average diameter as a function of molar charge. Strong TiO$_2$ ENPs destabilization occurs for Z >0.35 whereas, below this ratio, z-average diameter increase is linear indicating ENPs bridging.*

**Type I titration** - ζ potential value and binding heat of exchange as a function of SRHA over TiO$_2$ charge ratio are represented in Fig. 5a. For the first 15$^{th}$ SRHA addition (Z = [L]/[M] <0.35) the binding enthalpy slightly decreases in this domain, concomitantly with ζ potential which exhibits the same behavior due to the adsorption of negatively charged SRHA onto positively charged TiO$_2$ ENPs. It means that a large number of TiO$_2$ surface sites are still available for SRHA adsorption. Then a drastic decrease of the exchange energy is observed due to important surface charge modification leading to ENP charge inversion (-4.7 ± 0.4 mV) for Z = 0.4. After charge inversion has occurred, the energy of interaction is much lower due to electrostatic repulsions between the TiO$_2$-SRHA complexes and the titrant (SRHA) until the change of enthalpy is found constant for ζ potential values greater than -10 mV. Size evolution as a function of molar charge ratio (Fig. 5b) is in good agreement with our observations. For Z <0.4 the adsorption of SRHA on TiO$_2$ leads to the formation of agglomerates due to patch and bridging mechanisms with a linear increase of the z-average diameter value. Once charge neutralization and then charge inversion are achieved formation of large agglomerates (554 ± 28 nm) is observed. This sudden change of the agglomerate size corresponds to an important physical change (precipitation) after charge inversion as occurred. This change is also detected in the real time thermogram (Fig. 3) by a specific signature where important fluctuations of the heat flow are recorded (observed at the baseline level).

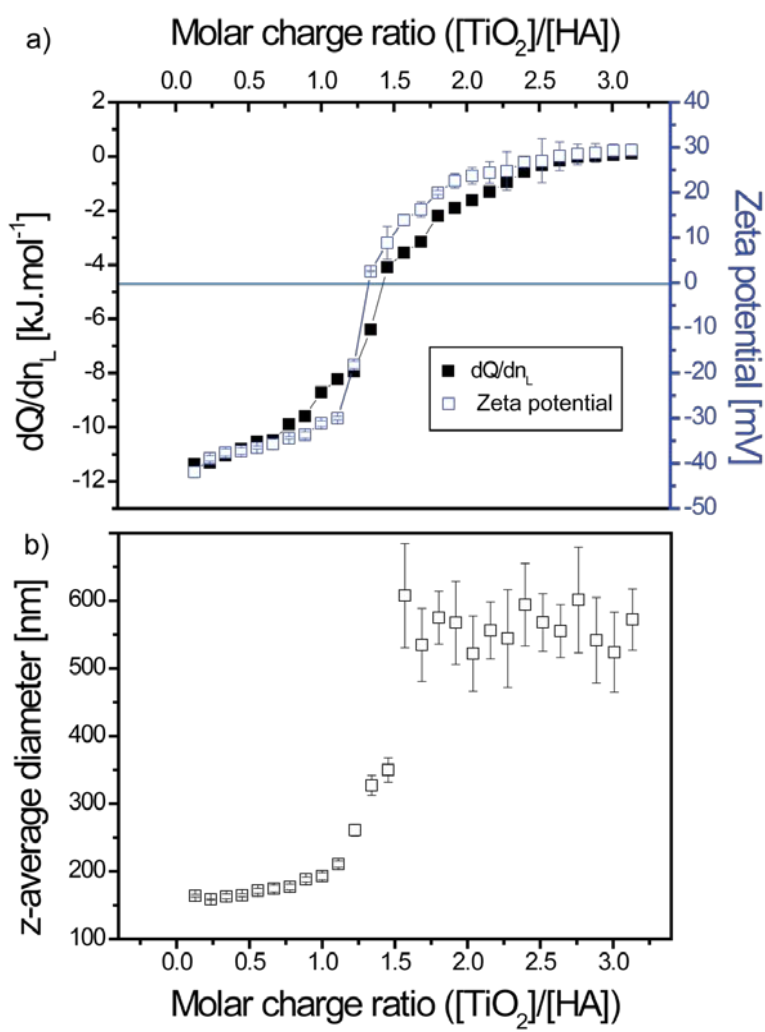

*Fig. 6 -* *a) Integrated heat data and ζ potential values as a function of TiO$_2$ over SRHA charge ratio for an SRHA 0.0375 mM titration with TiO$_2$ 0.7 g L$^{-1}$ at pH < pH$_{PCN,TiO_2}$. For a charge ratio ≥1.25, charge inversion is observed (ζ potential = +2.6 ± 0.2 mV) only weak interaction are observed. b) z-average diameter as a function of molar charge. Strong TiO$_2$*

*ENPs destabilization occurs for Z >1.5 whereas, below this ratio, z-average diameter increase is linear indicating ENPs bridging.*

**Type II titration** - $\zeta$ potential value and binding energy as a function of $TiO_2$ over SRHA charge ratio are represented in Fig. 6a. The influence of the charge modification on the heat exchange is clearly observed since the enthalpy of binding decrease is dependent on the surface charge and follows the decrease of $\zeta$ potential. For a charge ratio less than 1.25, $\zeta$ potential values are higher than -20 mV and the respective heat of exchange larger than -8 kJ mol$^{-1}$. Then further addition of $TiO_2$ induced the charge inversion of the $TiO_2$-SRHA complexes and thus an important decrease of the energy of association which becomes constant for $\zeta$ potential value higher than +25 mV. Z-average diameter value as a function molar charge ratio is shown in Fig. 6b. Before charge inversion the sizes of the complexes is slowly increasing due to patch and bridging mechanisms. Then the charge neutralization and further inversion leads to the formation of large agglomerates (562 ± 27 nm for Z >1.5).

For both titration types, a good agreement is found between the surface charge modification of $TiO_2$-SRHA complexes, the heat of exchange associated to the interaction processes as well as for the size evolution along the titrations. An interesting behavior for the interaction of $TiO_2$ ENPs in the presence of SRHA is that non negligible binding enthalpy is observed even after charge inversion (and particle precipitation) has occurred. It denotes the role of van der Waals interactions (especially for Type I) and change of conformation of SRHA. Such a behavior is not observed when linear natural polysaccharide (alginate) are considered [64]. This was due to the alginate chemical properties for which van der Waals interactions are not favored. Dynamic light scattering has also permitted to assign the real-time thermogram signature to an important precipitation domain in agreement with z-average diameter and $\zeta$ potential values.

# 4. Conclusion

The spontaneous association process between $TiO_2$ nanoparticles and Suwannee River humic acids has been shown to be dependent on concentration and mixing order. All interaction processes were found favorable from an enthalpic and entropic point of view and agglomeration was shown to be promoted by patch and bridging mechanisms.

This study shows the high potential of isothermal titration calorimetry (ITC) for the investigation of interactions between engineered nanoparticles (ENPs) and natural organic matter. Indeed ITC, especially when associated with light scattering techniques, not only allows the determination of important thermodynamic ($\Delta H$, $\Delta G$ and $\Delta S$) and reaction ($K_b$ and n) parameters but also permits a better understanding of the mechanism of interactions (and/or agglomeration) and the forces (hydrophobic, electrostatic) involved during association processes. ITC also gives quantitative and accurate information of the adsorption energies and hence potential reversibility of nanoparticle coating processes in various conditions. This novel technique, in environmental nanoscience area, also constitutes a potential promising instrumental method for the investigation of competitive sorption of environmental compounds (natural organic molecules and inorganic colloids) on ENPs. The major limitations of ITC concern the time consuming sample preparation and long analysis time but also the high concentrations needed (especially for the titrant) which can be a problem if working with costly materials or molecules being concentration conformational dependent. Nevertheless ITC is a non-destructive technique which permits to quantify the interfacial reactions (and the possible reversibility and stability of the association processes) which is primordial for a better holistic understanding of the transport and fate of (nano)particles in aquatic systems when exposed to a broad range of molecules of different abundance.

**Acknowledgments**
The authors are grateful to the financial support received from the Swiss National Foundation (200020_152847 and 200021_135240). The work leading to these results also received funding from the European Union Seventh Framework Programme (FP7/2007-20013) under agreement no NMP4-LA-2013-310451. L.V. also thanks the CNPq (Conselho Nacional de Desenvolvimento Científico e Tecnológico) 210694/2013-0 in Brazil for postdoctoral fellowship.

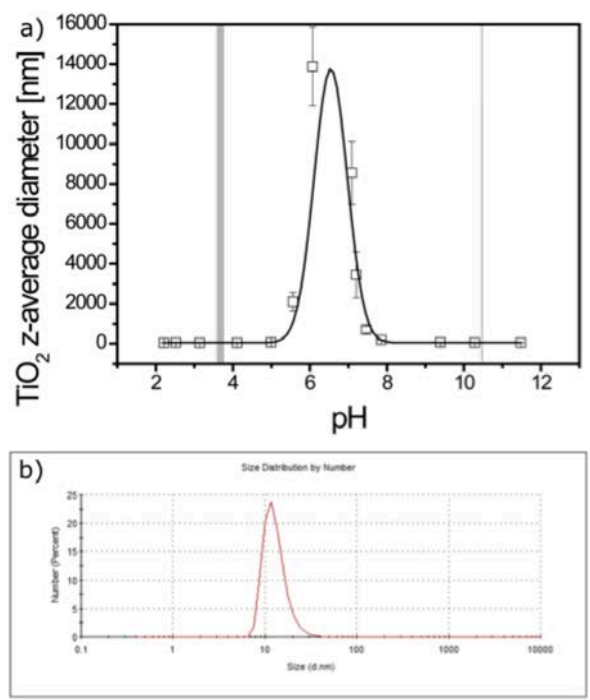

**Fig. S1 -** a) TiO$_2$ z-average diameter values as a function of pH. At pH 3.8 (large gray vertical line) the TiO$_2$ ENPs are dispersed with a z-average diameter value found equal to 47 ± 1 nm. At pH 10.4 (narrow gray line), the ENPs are also stable with diameter value equal to 53 ± 1 nm. b) TiO$_2$ number size distribution at pH 3.8 with mode value in agreement with manufacturer particle primary diameter [TiO$_2$] = 50 mg L$^{-1}$ and [NaCl] = 0.001 M.

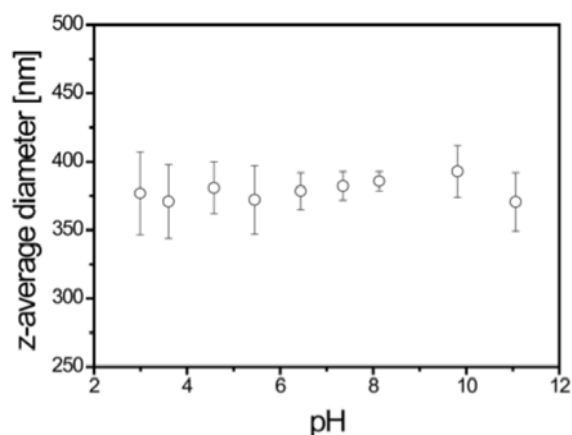

**Fig. S2 -** SRHA z-average diameter values as a function of pH. Z-average diameters are found constant over pH variation with mean value equal to 379 ± 19 nm. [SRHA] = 100 mg L$^{-1}$ and [NaCl] = 0.001 M.

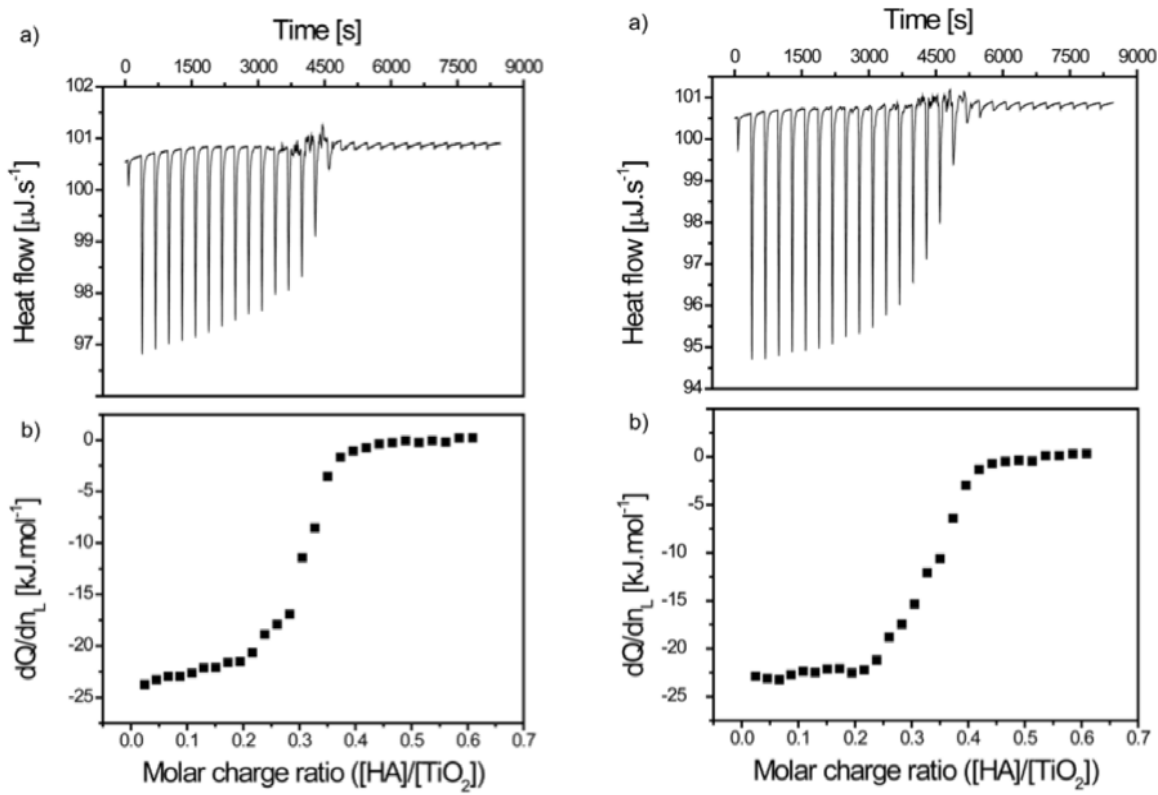

**Fig. S3 -** a) Real-time thermogram for $TiO_2$ 0.2 g L$^{-1}$ titration with SRHA 0.50 mM at pH < $pH_{PCN,TiO_2}$ at 298.15 K. Negative peaks indicate an exothermic reaction. b) Corresponding integrated heat data as a function of molar charge ratio.

**Fig. S4 -** a) Real-time thermogram for $TiO_2$ 0.3 g L$^{-1}$ titration with SRHA 0.75 mM at pH < $pH_{PCN,TiO_2}$ at 298.15 K. Negative peaks indicate an exothermic reaction. b) Corresponding integrated heat data as a function of molar charge ratio.

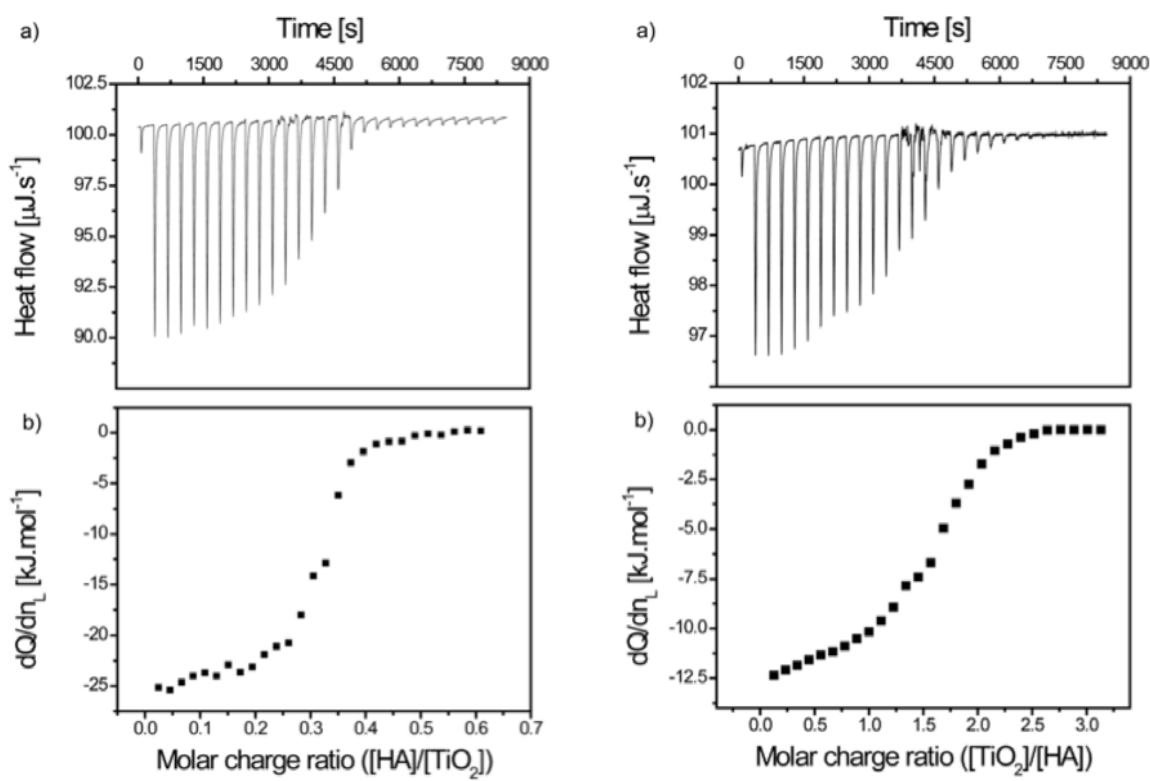

**Fig. S5 -** a) Real-time thermogram for TiO$_2$ 0.5 g L$^{-1}$ titration with SRHA 1.25 mM at pH < pH$_{PCN,TiO_2}$ at 298.15 K. Negative peaks indicate an exothermic reaction. b) Corresponding integrated heat data as a function of molar charge ratio.

**Fig. S6 -** a) Real-time thermogram for SRHA 0.075 mM titration with TiO$_2$ 1.4 g L$^{-1}$ at pH < pH$_{PCN,TiO_2}$ at 298.15 K. Negative peaks indicate an exothermic reaction. b) Corresponding integrated heat data as a function of molar charge ratio.

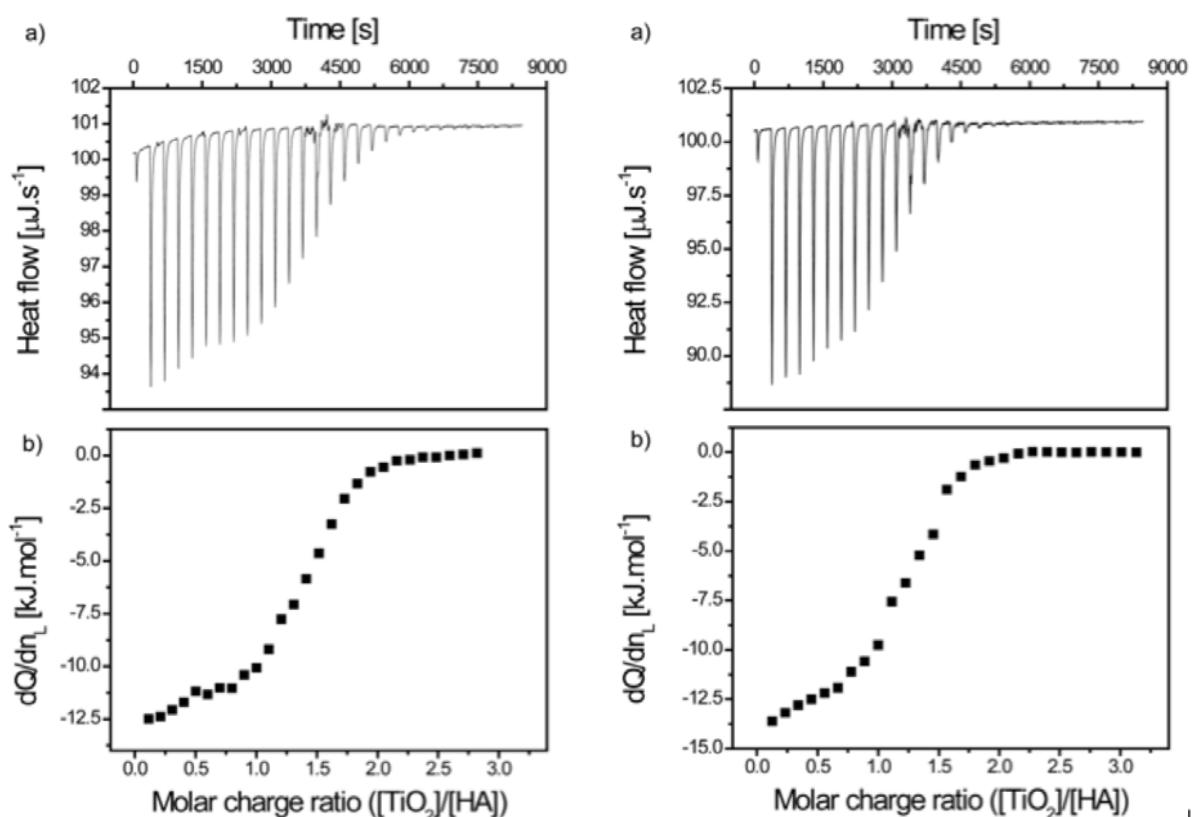

**Fig. S7** - a) Real-time thermogram for SRHA 0.1125 mM titration with TiO$_2$ 2.1 g L$^{-1}$ at pH < pH$_{PCN,TiO_2}$ at 298.15 K. Negative peaks indicate an exothermic reaction. b) Corresponding integrated heat data as a function of molar charge ratio.

**Fig. S8** - a) Real-time thermogram for SRHA 0.1875 mM titration with TiO$_2$ 3.5 g L$^{-1}$ at pH < pH$_{PCN,TiO_2}$ at 298.15 K. Negative peaks indicate an exothermic reaction. b) Corresponding integrated heat data as a function of molar charge ratio.

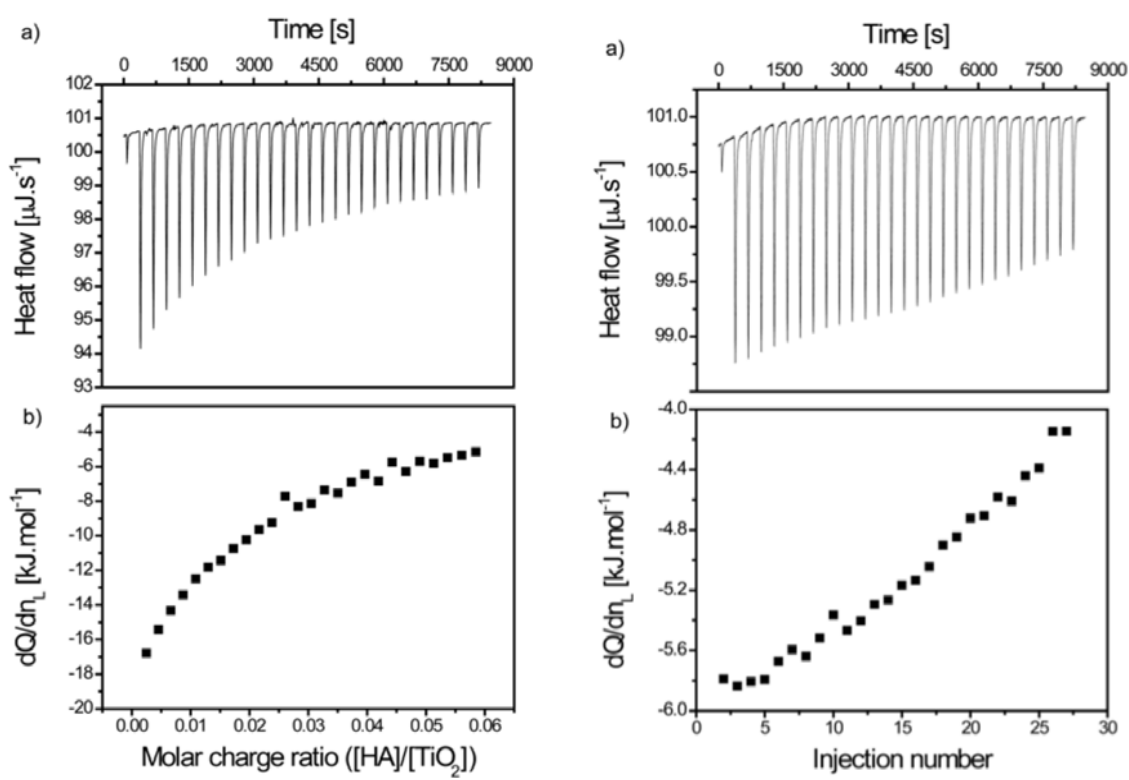

**Fig. S9** - a) Real-time thermogram for $TiO_2$ 5 g L$^{-1}$ titration with SRHA 1.25 mM at pH > $pH_{PCN,TiO_2}$ at 298.15 K. b) Corresponding integrated heat data as a function of molar charge ratio.

**Fig. S10** - a) Real-time thermogram for water titration with SRHA 1.25 mM at pH > $pH_{PCN,TiO_2}$ at 298.15 K. b) Corresponding integrated heat data as a function of injection number.